\newcommand*\diff{\mathop{}\!\mathrm{d}}

\documentclass[manuscript]{aastex6}
\usepackage{amsbsy}
\usepackage{amsfonts}
\usepackage{amssymb}
\usepackage{graphicx}
\usepackage{graphics}
\usepackage{bm}

\begin{document}

\title{Inferring the Coronal Density Irregularity from EUV Spectra}
\author{M. Hahn\altaffilmark{1} and D. W. Savin}
\affil{Columbia Astrophysics Laboratory, Columbia University, 550 West 120th Street, New York, NY 10027 USA}
\altaffiltext{1}{mhahn@astro.columbia.edu}

\date{\today}
\begin{abstract}
Understanding the density structure of the solar corona is important for modeling both coronal heating and the solar wind. Direct measurements are difficult because of line-of-sight integration and possible unresolved structures. We present a new method for quantifying such structure using density-sensitive EUV line intensities to derive a density irregularity parameter, a relative measure of the amount of structure along the line of sight. We also present a simple model to relate the inferred irregularities to physical quantities, such as the filling factor and density contrast. For quiet Sun regions and interplume regions of coronal holes, we find a density contrast of at least a factor of three to ten and corresponding filling factors of about 10--20\%. 
Our results are in rough agreement with other estimates of the density structures in these regions. The irregularity diagnostic provides a useful relative measure of unresolved structure in various regions of the corona. 
\end{abstract}

\keywords{techniques: spectroscopic, Sun: corona, Sun: atmosphere} 
	
\maketitle
	
\section{Introduction}\label{sec:intro}

Images of the solar corona show small-scale structures with transverse length scales down to the resolution limit of the instruments. This suggests unresolved structures at yet smaller length scales as well. The structures at scales near and below the resolution limits of current instruments are generally referred to as fine structure. Additional unresolved structure arises from the line-of-sight integration inherent in coronal observations, which obscures variations at length scales that might otherwise be resolvable. Characterizing all of this coronal density structure is necessary in order to resolve several problems in solar physics. 

One important issue is that of the observed Alfv\'enic wave dissipation in the corona \citep{Bemporad:ApJ:2012,Hahn:ApJ:2012,Hahn:ApJ:2013,Hahn:ApJ:2014}. These waves appear to be damped closer to the Sun than predicted by simple models \citep[see e.g.,][]{Cranmer:SSR:2002}. This surprisingly rapid damping may be explained by inhomogeneities in the corona that increase the damping rate, such as by reflection off of density fluctuations along the magnetic field that increase the reflection rate and drive turbulence \citep{vanBallegooijen:2016} or by phase mixing between inhomogeneous layers transverse to the mean magnetic field direction \citep[e.g.,][]{Heyvaerts:AA:1983, DeMoortel:AA:2000}. A related issue concerning waves is that information about the density structure is required in order to accurately determine the wave energy flux from observations \citep{Goossens:ApJ:2013}. 

Another area in which density structuring is important is for understanding solar wind composition, which is an important tool for determining the source regions of the solar wind. The charge state distribution of the solar wind becomes frozen in close to the Sun, as the density decreases and collisions become infrequent. Modeling the evolution of solar wind charge states therefore requires accurate knowledge of the density structure in the sources of the solar wind. If, for example, there are structures with significantly higher density than the average density inferred from spectroscopy, then there would be more collisions in those structures and the charge states would continue evolving to larger heights than they do in less dense regions. 

Measurements of the density structure of the corona are limited by the spatial resolution of the instruments, and, more fundamentally, by the line-of-sight averaging inherent in observations. Occasionally, there are opportunities to quantify small-scale structure using unique observations. Images of the corona from solar eclipses processed to highlight fine structures reveal numerous filaments with transverse length scales of 1--5~Mm and density contrasts of about a factor of two \citep{November:ApJ:1996}. \citet{Raymond:ApJ:2014} were able to infer properties of coronal structure by studying the ionization of gas from a sun-grazing comet, finding density variations over length scales of several megameters and density contrasts of more than a factor of six. Another approach that has been proposed is to use observations of resonant damping of kink magnetohydrodynamic waves to infer the density structure \citep{Arregui:ApJ:2013}.

Routine measurements of density structure can be performed using spectroscopy, but existing diagnostics have several limitations. The most detailed information about density structure that could be obtained from spectroscopy would be to infer the full emission measure distribution differential in both temperature and density \citep{Almleaky:AA:1989,Brown:AA:1991}. However, it is currently impractical to perform such an analysis because spectroscopic density diagnostic lines have similar dependences on density \citep[see][]{Judge:ApJ:1997} and because there are large uncertainties in the atomic data \citep[e.g.,][]{Young:AA:2009}.  

\citet{Orrall:ApJ:1990} proposed a method for quantifying unresolved density structure that combined extreme ultraviolet (EUV) spectra with coronagraph measurements of the K-corona polarized brightness $pB$. Resonance lines observed in the EUV are sensitive to the electron density squared, $n_{\mathrm{e}}^2$, while $pB$ is sensitive to $n_{\mathrm{e}}$. Thus, the ratio of the measured EUV intensity to $pB^2$ is proportional to the ratio of the average of the density squared to the square of the average density, 
$\overline{n_{\mathrm{e}}^2}/(\overline{n_{\mathrm{e}}})^2$, 
which \citeauthor{Orrall:ApJ:1990} refer to as the coronal density irregularity.

Here, we implement a diagnostic similar to that of \citet{Orrall:ApJ:1990}, but based solely on EUV spectra. We apply this diagnostic to observations of coronal hole and quiet Sun regions obtained by the EUV Imaging Spectrometer \citep[EIS;][]{Culhane:SolPhys:2007} on \textit{Hinode}. Typically, EUV emission lines are used as density diagnostics by taking the intensity ratio of two lines, one of which is density sensitive and the other which is not. The EIS data contain several of these density-diagnostic line pairs. Our approach is to first, assume that the corona has a smooth, large-scale density distribution, which we take to be spherically symmetric with a scale-height falloff. We calculate the radial density profile that would result in the observed line intensity ratio. Then, based on that density profile, we calculate the absolute intensities for each line in the diagnostic pair. Finally, we determine the ratio of the observed intensities to the modeled absolute intensities. We define this ratio as the irregularity parameter, which characterizes the density structure by quantifying the degree to which the actual corona differs from the assumed model. With additional modeling, we can interpret the irregularity factor in terms of more physical quantities, such as filling factor and density contrast.

Our method offers several advantages compared to other techniques for quantifying the coronal structure. Since we rely only on UV spectra, all of the needed data can be obtained from the same instrument, thereby eliminating cross-calibration errors that arise when using data from different instruments. Also, because the emission line pairs can come from different ion species, temperature information is included in the spectral data. This allows us to distinguish between structures at different temperatures and thereby incorporate aspects of a full emission measure analysis. 

The rest of this paper is organized as follows: Section~\ref{sec:method} describes the analysis procedure we use to determine the density irregularity from the EUV intensities. In Section~\ref{sec:obs} we apply this diagnostic to observations of coronal holes and quiet Sun regions and report the inferred irregularities. We interpret these irregularities in Section~\ref{sec:interpret} in the context of a slab model in order to estimate physical properties of the density structure in the observations. Section~\ref{sec:discuss} concludes. 

\section{Inferring the Irregularity}\label{sec:method}

The procedure for inferring the irregularity is as follows: First, we measure the intensities of emission lines that form a density-diagnostic line pair. Next, we use a spherically-symmetric scale-height-falloff density distribution to model the observed intensity ratio. By integrating the intensities along a line of sight through the model, we determine the density required at the base of a spherically symmetric corona that would lead to the same intensity ratio as is observed. Finally, we calculate the absolute line intensities from each of the two lines in the diagnostic pair and compare them to the observed intensities. The ratio of the observed intensities to the model intensities gives the irregularity, $X$. 
The analysis is explained in more detail below. 

\subsection{Density Diagnostic Lines}\label{subsec:density}

The intensity of a spectral line emitted by a transition from level $j$ to level $i$ for a given ion can be expressed in terms of the emissivity $\epsilon_{ij}$ as 
\begin{equation}
I_{ji} = \frac{1}{4\pi} \int \epsilon_{ij}(T, n_{\mathrm{e}}) \diff{h},
\label{eq:int_emiss}
\end{equation}
where $T$ and $n_{\mathrm{e}}$ are the electron temperature and density at the source of the emission and $\diff{h}$ lies along the line of sight. The emissivity is defined as 
\begin{equation}
\epsilon_{ij}= \frac{hc}{\lambda_{ij}} n_{j} A_{ji}. 
\label{emissdef}
\end{equation}
Here, $h$ is the Planck constant, $c$ is the speed of light,  $\lambda_{ij}$ is the wavelength of the transition so that $hc/\lambda_{ij}$ is the transition energy, $n_{j}$ is the population number density of the upper level $j$ that produces the emission line, and $A_{ji}$ is the Einstein A-rate for the transition. It is often useful to express the emissivity as a multiple of various ratios, thereby defining the contribution function $G(T,n_{\mathrm{e}})$ as 
\begin{equation}
G_{ji}(T, n_{\mathrm{e}}) = \frac{hc}{\lambda_{ij}} \frac{n_{j}(\mathrm{X}^{+q})}{n(\mathrm{X}^{+q})}\frac{n(\mathrm{X}^{+q})}{n(\mathrm{X})}\frac{n(\mathrm{X})}{n(\mathrm{H})} \frac{n(\mathrm{H})}{n_{\mathrm{e}}}\frac{A_{ji}}{n_{\mathrm{e}}},
\label{eq:Gdef}
\end{equation}
where $n_{j}(\mathrm X^{+q})/n(\mathrm{X}^{+q})$ is the relative population 
of the upper level $j$ for charge state $\mathrm{X}^{+q}$, 
$n(\mathrm{X}^{+q})/n(\mathrm X)$ is the relative abundance for charge state 
$q$ of element $\mathrm{X}$, $n(\mathrm{X})/n(\mathrm{H})$ is the abundance 
of $\mathrm{X}$ relative to hydrogen, and $n(\mathrm{H})/n_{\mathrm{e}}$ is the 
hydrogen density relative to that of free electrons. In this case, 
\begin{equation}
I_{ji} = \frac{1}{4\pi} \int G(T, n_{\mathrm{e}}) n_{\mathrm{e}}^{2} dh. 
\label{eq:intensity}
\end{equation}
In order to calculate the emissivities and contribution functions, we use the CHIANTI database, which tabulates the necessary atomic data \citep{Dere:AA:1997,DelZanna:AA:2015}. 

The electron density $n_{\mathrm{e}}$ is inferred spectroscopically using the intensity ratio of two spectral lines. These lines are usually chosen to be from the same ion so that the dependence on charge states and elemental abundances cancel.
The physics behind density-sensitive line ratios are explained in detail in \citet[][]{Phillips:Book}.

\begin{figure}[h]
	\centering \includegraphics[width=0.9\textwidth]{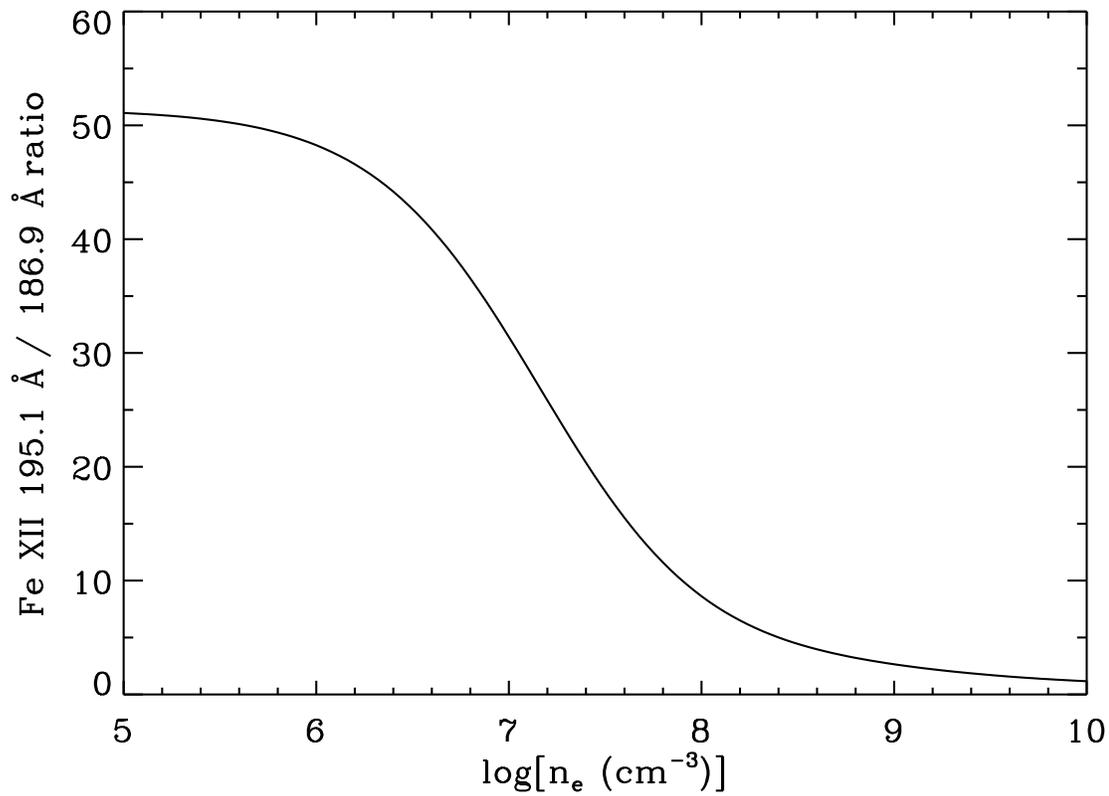}
	\caption{\label{fig:denex} Intensity ratio versus the log of the electron density for the Fe~\textsc{xii} 186.9~\AA\ $/ 195.1$~\AA\ lines.
	}
\end{figure}

Table~\ref{table:densitylines} lists the transitions we use here as density diagnostics. Figure~\ref{fig:denex} shows the theoretical intensity ratio for the Fe~\textsc{xii} $195.1$~\AA$/186.9$~\AA\ diagnostic. Note that the intensity ratio varies in a complex way as a function of $n_{\mathrm{e}}$. In particular, there is a region of moderate densities at which the ratio varies strongly as a function of density, but at very high and very low densities, the ratio tends to saturate. 

For a nonuniform plasma, the density inferred from a line ratio represents an average density along the line of sight. This average is weighted by the sensitivity of the diagnostic to the various densities sampled. Consequently, the inferred density differs from the true density of any actual structure within the observed region \citep{Doschek:ApJ:1984}. 

\subsection{Spherically Symmetric Structure}\label{subsec:spherical}

In order to characterize the fine-scale density structure of the corona, it is first necessary to take into account the effects of density structure at the largest relevant scale. It is commonly observed that the density decreases radially with height in the corona. Here, we assume that this radial density fall-off is spherically symmetric. This assumption ignores medium-scale structures along the line of sight, such as might occur if the line of sight passes through a coronal hole with some quiet Sun in the foreground or background. For such cases, we expect our assumption to be reasonable provided that the path length through the structure, meaning in this example the coronal hole, is long compared to the scale height. If this criterion is satisfied, then the density along the line of sight through the secondary structure is low and its contribution to the total emission is negligible. 

At low heights, where the solar wind velocity is small, the density in the corona is well-described by an exponential fall-off \citep[e.g.,][]{Landi:ApJ:2006,Wilhelm:AAR:2011, Hahn:ApJ:2013}, 
\begin{equation}
n_{\mathrm{e}}(r)=n_{\mathrm{e}}(R_{\sun})\exp \left[-(r-1)/(rH)\right], 
\label{eq:radialdensity}
\end{equation}
where $r$ is the radius measured from Sun center in units of $R_{\sun}$, $n_{\mathrm{e}}(R_{\sun})$ is the density at 1~$R_{\sun}$, and $H$ is the scale height. The scale height is given, in units of $R_{\sun}$, by 
\begin{equation}
H=\frac{1}{R_{\sun}}\frac{ k_{\mathrm{B}} T}{\mu g_{\sun}},
\label{eq:scaleheight}
\end{equation}
where $k_{\mathrm{B}}$ is the Boltzmann constant, $\mu \approx 0.61 m_{\mathrm{p}}$ is the mean particle mass, $m_{\mathrm{p}}$ is the proton mass, and $g_{\sun}$ is the solar surface gravity.

In our analysis, described in Section~\ref{sec:obs}, we determine the temperature $T$ from a differential emission measure analysis and then calculate the corresponding $H$ to use for the structure analysis. We also limit our analysis to low heights, around 1.05~$R_{\sun}$, where the solar wind is slow so that non-hydrostatic effects are expected to be small.

A spherically symmetric function can be integrated along the line of sight by changing variables to form the Abel transform. Suppose $y$ is the projected height above the solar disk of the observation. That is, $y$ corresponds to the radius at which the line of sight passes closest to the solar surface. Then the integral of a radial function $f(r)$ along the line of sight is 
\begin{equation}
\int_{y}^{\infty} \frac{f(r)r}{\sqrt{r^2 - y^2}} \diff{r}. 
\label{eq:Abel}
\end{equation}
\citet{Orrall:ApJ:1990} show that the scale-height expression given by Equation~(\ref{eq:radialdensity}) for $n_{\mathrm{e}}(r)$ or $n_{\mathrm{e}}(r)^2$ can be integrated analytically in terms of modified Bessel functions. 

Using the density-dependent emissivities from CHIANTI, we can model the observed intensities expected from an isothermal spherically symmetric corona at a given height. The emissivities inherits spherical symmetry from their dependence on $n_{\mathrm{e}}(r)$. But, although each $\epsilon_{ij}(r)$ is spherically symmetric, their functional dependence on $n_{\mathrm{e}}(r)$ is complex and so we perform the integration numerically. We use a semi-open integration formula to account for the divergence of the integral at the lower limit and set the step sizes to $10^{-3}H$ \citep{Press:Book}. The integral is truncated at a maximum radius of $y+3Hy$. Tests with smaller step sizes and higher upper limits indicate that with these parameters, the integrals are accurate to better than 1\%. 

The large-scale spherically symmetric part of the density structure itself has several implications for interpreting observations. First, the density derived from intensity ratio diagnostics at some nominal height $y$ is significantly different from the density at the corresponding radius, where $r=y$. This is because line-of-sight averaging reduces the inferred density. 

\begin{figure}[h]
	\centering \includegraphics[width=0.9\textwidth]{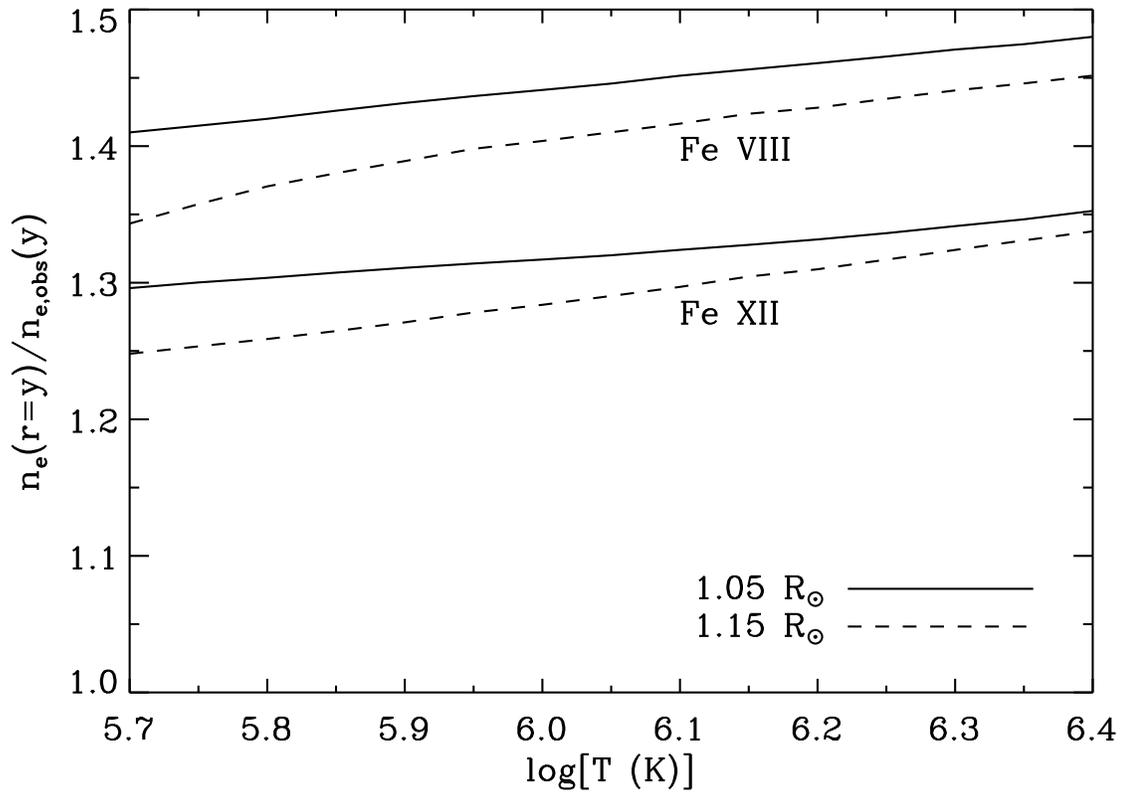}
	\caption{\label{fig:sphereden} Ratio of the input model density $n_{\mathrm{e}}= 1\times 10^{9}$~$\mathrm{cm^{-3}}$ relative to the corresponding density inferred from the synthetic observation $n_{\mathrm{e,obs}}$, which is smaller due to the line-of-sight averaging. The ratios are plotted as a function of $T$. Solid curves show the results for $y=1.05$~$R_{\sun}$ and dashed curves for $y=1.15$~$R_{\sun}$. The upper two curves are for the Fe~\textsc{viii} diagnostic and the lower two are for the Fe~\textsc{xii} diagnostic.
	}
\end{figure}

Figure~\ref{fig:sphereden} illustrates the effect of the line of sight on the inferred density. For this figure a base density $n_{\mathrm{e}}(R_{\sun})=1\times10^{9}$~cm$^{-3}$ was assumed. The intensity for the Fe~\textsc{viii} and Fe~\textsc{xii} density diagnostics were then calculated along rays at various heights $y$. We then took the ratio of the model intensities within each pair and found the corresponding inferred density, $n_{\mathrm{e,obs}}$ for our synthetic observation. One consequence of the line-of-sight averaging that we find is that the actual radial density in the model at $r=y$ is greater then the density inferred from the synthetic observation by a factor of $n_{\mathrm{e}}(r)/n_{\mathrm{e,obs}} \sim 1.3$. This ratio depends on temperature through the dependence on the scale height $H$, which is proportional to $T$. This influence, however, is weak because increasing $H$ both makes the density more uniform and also increases the effective path length through the structure. Some temperature dependence also arises from the temperature sensitivity of the emissivity. Additionally, the ratio depends on height, decreasing towards unity at larger heights. This is because eventually the density becomes so low that the line ratio is no longer density sensitive.

\begin{figure}[h]
	\centering \includegraphics[width=0.9\textwidth]{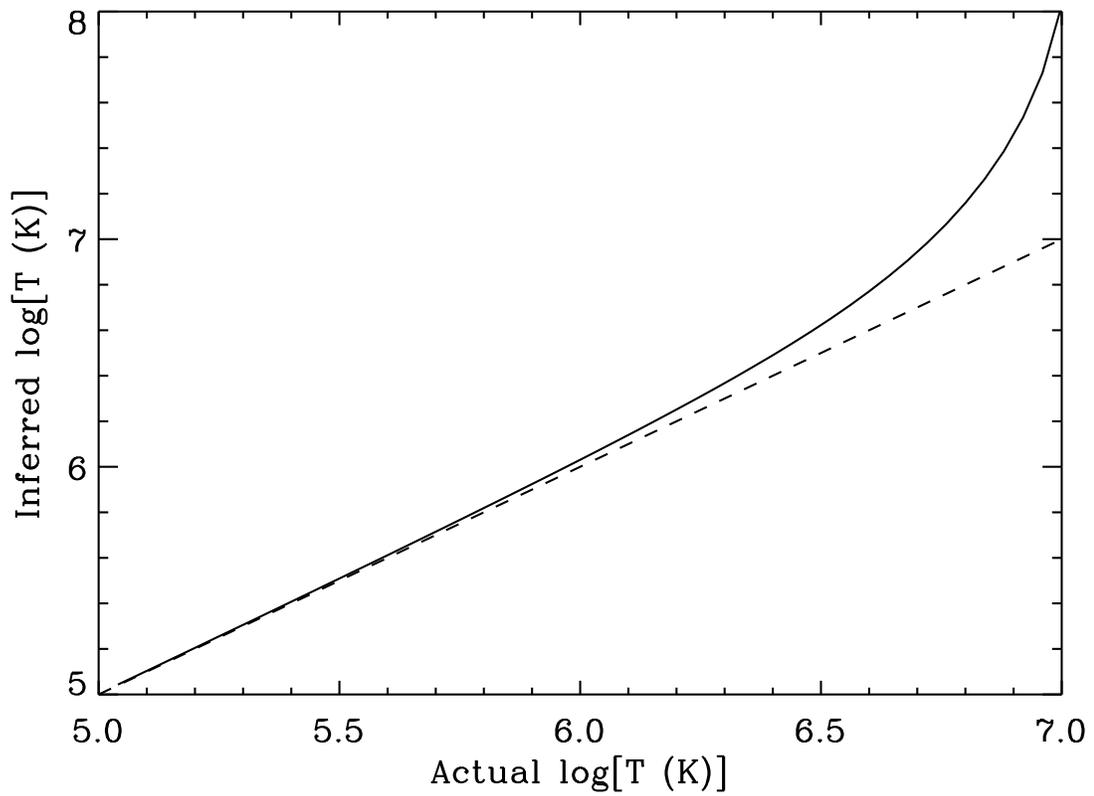}
	\caption{\label{fig:Tscale} The temperature inferred by fitting the emission measure versus height profile to a scale-height function from a model observation (solid curve) versus the temperature input in the model (dashed curve). The inferred temperatures are systematically greater than the actual temperatures.
}
\end{figure}

A second consequence of the line-of-sight integration is that the intensity decreases more slowly with nominal height $y$ than would be expected from a scale-height falloff. Thus, the apparent scale height is larger than the true scale height and analyses that derive $T$ from the apparent $H$ are systematically biased towards higher temperatures than the actual one. To quantify this effect, we have calculated the emission measure EM along the line of sight, defined as
\begin{equation}
\mathrm{EM}=\int n_{\mathrm{e}}^2 \diff{h}. 
\label{eq:emdef}
\end{equation}
For lines that are not density sensitive, the observed intensity is proportional to the EM, $I=\mathrm{EM} \, G(T)/(4\pi)$. We calculated the $\mathrm{EM}(y)$ for various input temperatures $T$. We then fit the results to a scale-height function similar to Equation~\ref{eq:radialdensity}, but using the projected nominal height $y$ in place of the actual radius $r$. From the fit we determined the scale-height, which gives the inferred scale-height temperature. Figure~\ref{fig:Tscale} shows the inferred $T$ compared to the actual input $T$. We find that the inferred temperature is always greater than the input temperature, and that the systematic error increases with temperature, being about 10\% at 1~MK and growing to greater than 50\% above 4~MK. 

This effect might be one reason why scale-height temperatures inferred for many coronal structures often are greater than what is found using other types of temperature diagnostics  \citep[e.g.,][]{Landi:ApJ:2006,Hahn:ApJ:2013}. Other sources for the discrepancy have been proposed, including that the observed region is not in hydrostatic equilibrium, multiple structures along the line of sight, multithermality, temperature variations with height, and gravitational settling. These factors can also contribute to the discrepancies, in addition to the line-of-sight effect described above.

\subsection{Coronal Density Irregularity}\label{subsec:finestruct}

The fine density structure can be characterized after accounting for the spherically symmetric large-scale structure. We begin with two observed line intensities, $I_{1,\mathrm{obs}}$ and $I_{2,\mathrm{obs}}$, from a density diagnostic pair. The ratio of these lines depends on the line-of-sight averaged density. Using the spherically symmetric model, we perform a least squares fit to the observed averaged density to find the parameter $n_{\mathrm{e}}(R_{\sun})$ in Equation~(\ref{eq:radialdensity}). That is, we find the value of $n_{\mathrm{e}}(R_{\sun})$, which produces the same intensity ratio as what is observed. The absolute intensities may, however, be different.

We define the density irregularity parameter $X$ to be the ratio of the observed intensity of either line to the corresponding model intensity, where the modeled density-sensitive line-pair ratio matches the observed one. That is, 
\begin{equation}
X \equiv \frac{I_{1,\mathrm{obs}}}{I_{1,\mathrm{mod}}} = \frac{I_{2,\mathrm{obs}}}{I_{2,\mathrm{mod}}},
\label{eq:xdef}
\end{equation}
where the equivalence of $X$ based on either of the two absolute intensities follows from the requirement that the model have $I_{1,\mathrm{mod}}/I_{2,\mathrm{mod}} = I_{1,\mathrm{obs}}/I_{2,\mathrm{obs}}$.  

The quantity $X$ reflects the deviation of the observed structure from the assumed model. Thus, if the observation has a density distribution matching that assumed, then we would find $X = 1$. The degree to which $X$ differs from unity indicates how much more structured the observation is relative to the model.

Although $X$ is a measure of unresolved structure, its relation to physical quantities is model-dependent. For the same structure, different density diagnostics can produce different values of $X$ because each line pair can have a different sensitivity to density. In particular, the value of $n_{\mathrm{e}}(R_{\sun})$ found as an intermediate step in the analysis is not necessarily the real density at the base of the corona, but only the density that a spherically symmetric hydrostatic corona would need to have if it were to produce the observed line intensity ratio. Thus, irregularity measurements from different diagnostics are not directly comparable. It is, however, valid to compare relative measurements of $X$ from the same pair of lines applied to different observations. 

In order to relate $X$ to physical properties of the corona, some model of the coronal density structure is needed. In Section~\ref{sec:interpret} we interpret our observational results in terms of a slab model. Although this is somewhat simplistic, it is the implicit context within which existing measurements of coronal fine structure are already interpreted when referring to filling factors and density contrasts.

Throughout our irregularity analysis we assume an isothermal corona where the radial variation is hydrostatic. These assumptions are not necessary for the irregularity analysis. It would be straightforward to incorporate non-hydrostatic radial variation or a large scale structure that is not spherically symmetric. The isothermal assumption could be relaxed by allowing for a radial temperature gradient. As discussed above, our assumption of spherical symmetry is reasonable if the largest relevant length in the observation is much greater than the scale height and it is also analytically tractable. However, a possible refinement of our analysis would be to use a large scale numerical reconstruction of the corona to represent the large-scale density structure and then apply the irregularity analysis to quantify the fine structure relative to that model.

\section{Observations and Results}\label{sec:obs}

We have determined the coronal density irregularity using several line pairs in three EIS observations of coronal holes and quiet Sun regions. Here we describe those parts of the analysis that are common to all three observations, while aspects unique to particular datasets are described in Sections~\ref{subsec:hole} and \ref{subsec:qs}. 

For each observation the data were prepared using the standard EIS processing routines to clean the data of spikes and warm pixels, calibrate the intensities, and correct for wavelength-dependent spatial offsets. We also spatially binned the data by first aligning all the data to the same wavelength scale before averaging the intensities. The bin sizes we used are different for each observation.

The EIS absolute calibration has been drifting since the beginning of the \textit{Hinode} mission and several calibration schemes have been proposed. We have performed the analysis mainly using the calibration proposed by \citet{Warren:ApJS:2014}, but have also tested the effects of using the alternate calibration proposed by \citet{DelZanna:AA:2013}. The sensitivity of our results to the absolute calibration is discussed in more detail in Section~\ref{subsec:uncertainties}. 

We extracted line intensities by fitting Gaussian functions to the spectra \citep[see e.g.,][]{Hahn:ApJ:2012}. In addition to the lines given in Table~\ref{table:densitylines}, we also analyzed other prominent lines in the EIS spectrum, which we used for the temperature analysis. Line lists are given in our previous papers for coronal holes \citep{Hahn:ApJ:2013} and quiet Sun regions \citep{Hahn:ApJ:2014}. Most of the lines we used were unblended, or are self-blends that arise from the same ion species, where the analysis can be performed by summing the emissivities of the individual contributors to the blend. The main exception is the Si~\textsc{vii} line, which is used in the irregularity analysis.

The Si~\textsc{vii} 274.18~\AA\ line is blended with an Fe~\textsc{xiv} line at 274.20~\AA. In principle, the contribution of the Fe~\textsc{xiv} line to the blend could be subtracted by measuring the Fe~\textsc{xiv} line at 289.15~\AA\, which has a fixed branching ratio with the blended Fe~\textsc{xiv}. In practice, the longer wavelength line is difficult to measure, because EIS has a small effective area at that wavelength. Instead, we ignored this line for the DEM portion of our analysis. But, we then used the derived DEM to estimate the fraction of the emission from the Fe~\textsc{xiv} line and subtracted this from the measured 274.2~\AA\ line intensity. Since the Fe~\textsc{xiv} intensity makes up the majority of the blend, the uncertainties on the Si~\textsc{vii} line intensity are large and the diagnostic was not useful for some observations.

For each position where we measured the density irregularity, we also performed a DEM analysis for the temperature. The DEM describes the amount of material along the line of sight as a function of temperature. The DEM, $\phi(T)$, is related to measured intensities through
\begin{equation}
I_{ji}=\frac{1}{4\pi}\int G(T) \phi(T) \diff{T}.
\label{eq:demdef}
\end{equation}
Here we neglect the density-dependence of $G(T)$ by using lines that are not density sensitive for the DEM analysis. In order to test for possible systematic errors in the DEM analysis caused by the inversion method, we carried out the analysis using several techniques, including the iterative method of \citet{Landi:AA:1997}, the regularized inversion method of \citet{Hannah:AA:2012} and the least squares fitting method using Gaussians described in \citet{Hahn:ApJ:2014}. 

The DEM-averaged formation temperature $T_{t}$, is useful because it indicates the temperature within the plasma that dominates the emission in a particular line. For any line, labeled by $ij$, this quantity is given by
\begin{equation}
\log T_{t} = \frac{\int G_{ji}(T) \phi(T) \log(T) \diff{T}}{\int G_{ji}(T)\phi(T)\diff{T}}. 
\label{eq:Ttdef}
\end{equation}

For each observation, we focused on positions at the same height of about $y=1.05$~$R_{\sun}$. This height was chosen because it is expected to be high enough to mitigate contributions from chromospheric structures and limb brightening. But it is also low enough that instrument scattered light is neglibile and the emission lines are bright so that they can be measured accurately. 

The irregularity analysis was performed for each pair of diagnostic lines that had sufficient data. In some cases, a line pair could not be used due to low signal. For example, the Fe~\textsc{viii} lines are dim in the quiet Sun, while the Fe~\textsc{xiii} lines are weak in the coronal holes. Uncertainties in the measured line intensities were propagated into the irregularity analysis by using a Monte-Carlo error analysis. That is, we generated a Gaussian distribution of random numbers with the same mean as the measured intensities and the same standard deviation as the corresponding uncertainty. We then re-calculated the irregularity for each pair of random intensities. We report the standard deviation of the resulting $X$ values as the uncertainty. 

We also performed the irregularity analysis as a function of temperature. The absolute intensities can depend strongly on temperature, especially where the temperature of the emitting region is away from the peak formation temperature of the ion emitting the line. When we report a value of $X$, the temperature at which it is valid is also given. This is generally either the peak temperature in the DEM or the DEM-weighted average formation temperature. Some examples of plots of $X$ versus the assumed temperature are given below, which illustrate the sensitivity of our results to the temperature analysis.

Other sources of systematic error exist. These include the absolute calibration of EIS and atomic data uncertainties. We consider their possible effects in Section~\ref{subsec:uncertainties}. 

\subsection{Coronal Holes}\label{subsec:hole}

One of the coronal hole observations we studied was made on 2009 April 23 using the $2^{\prime\prime}$-wide slit of EIS. This observation actually consisted of five slit positions centered with respect to the solar meridian at, $-44.5^{\prime\prime}$, $-14.5^{\prime\prime}$, $15.5^{\prime\prime}$, $45.4^{\prime\prime}$ and $105.6^{\prime\prime}$. In each case the vertical axis of the 512$^{\prime\prime}$ long slit was centered at about $-1140^{\prime\prime}$. Each of these pointings observed primarily interplume coronal hole plasma. These five pointings were summed by averaging the data radially to form a single composite pointing. The details of this averaging are described by \citet{Hahn:ApJ:2012}. Here, we have futher binned the data vertically in 8 pixel bins to increase the signal to noise. We label this observation CH-A. 

\begin{figure}[h]
	\centering \includegraphics[width=0.9\textwidth]{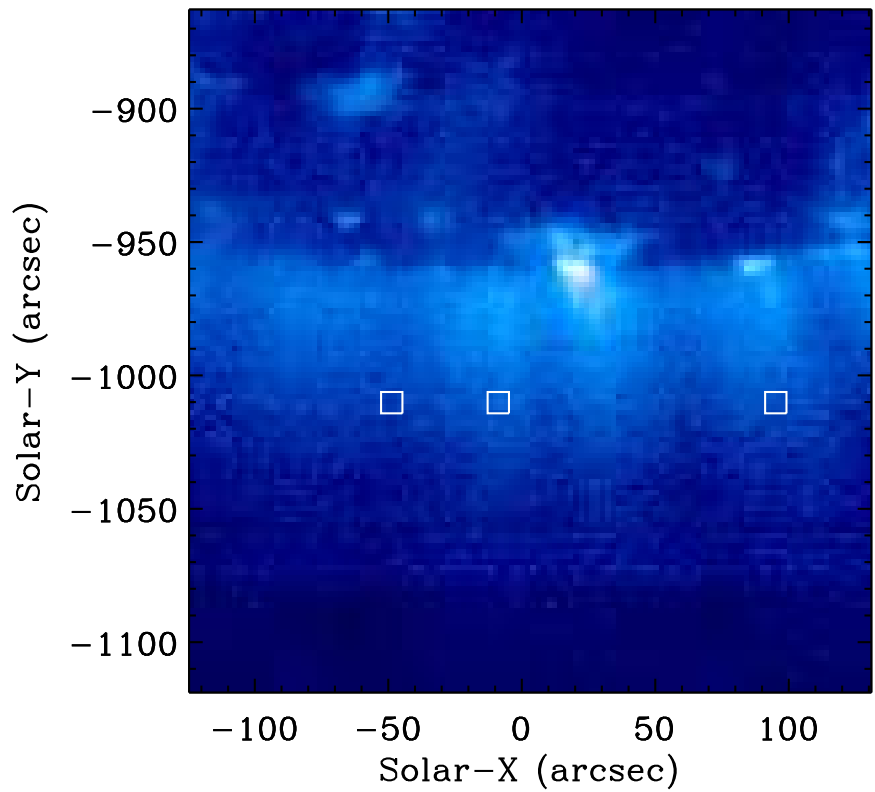}
	\caption{\label{fig:pipimage} Image of the Fe~\textsc{ix} 197.86~\AA\ line intensity for the coronal hole observation CH-B. Several bright plume and dark interplume regions are clearly visible. The particular regions that we studied are outlined by boxes and are located at Solar-X coordinates of about $-50^{\prime\prime}$, $-10^{\prime\prime}$, and $90^{\prime\prime}$. The background image here has $2^{\prime\prime} \times 2^{\prime\prime}$ pixels, but for our analysis we used a larger binning of $8^{\prime\prime} \times 8^{\prime\prime}$, as indicated by the box sizes. 
}
\end{figure}

The other coronal hole observation, which we label CH-B, is a raster observation made on 2007 March 13 made with the $1^{\prime\prime}$ EIS slit that spanned from $-124.13^{\prime\prime}$ to $130.57^{\prime\prime}$ horizontally with respect to the central meridian of the Sun. The slit was centered vertically at $-972.43^{\prime\prime}$. This observation was studied previously by \citet{Guennou:ApJ:2015} and more details can be found there. The data contain several clear interplume and plume regions. For our analysis we focused on the interplume region at $\approx -50^{\prime\prime}$ and plume regions at $\approx -10^{\prime\prime}$ and $\approx 90^{\prime\prime}$. Figure~\ref{fig:pipimage} shows the entire EIS observation as seen in the Fe~\textsc{ix} 197.86~\AA\ line. The particular regions we studied are outlined by the boxes, which have the same size as the $8^{\prime\prime}$ by $8^{\prime\prime}$ pixel bins that we used for CH-A.

\begin{figure}[h]
	\centering \includegraphics[width=0.9\textwidth]{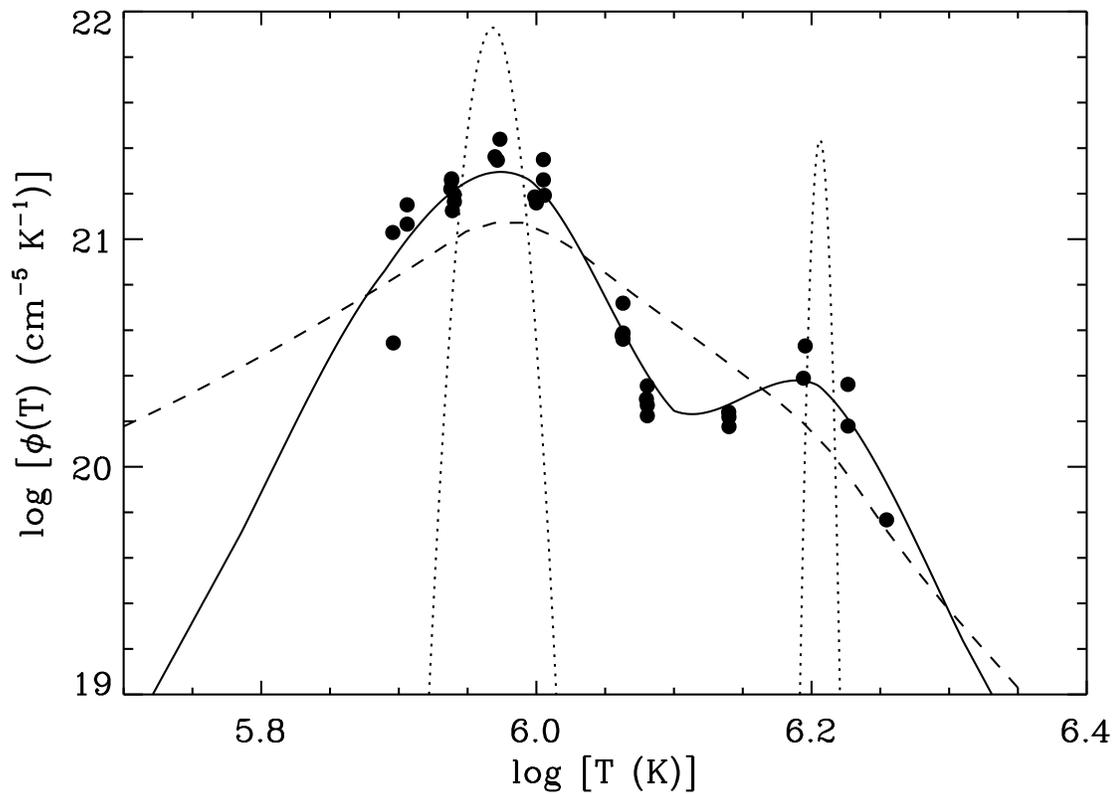}
	\caption{\label{fig:chdem} DEM at 1.05~$R_{\sun}$ for the CH-A observation of an interplume coronal hole region. The DEM results derived from several techniques: the iterative method (solid curve), the regularized inversion (dashed curve), and fit to two Gaussians (dotted curve). The filled circles show the ratio of the measured intensity to the DEM-derived intensity at $\log T_{t}$ for the iterative DEM analysis, and their scatter about the DEM curve is an indication of the quality the result. 
}
\end{figure}

Table~\ref{table:irreg} summarizes our irregularity results for these observations as well as the quiet Sun observation described below, in Section~\ref{subsec:qs}. In each case, we specify the temperatures that are most likely to be relevant. For example, the DEM for the CH-A observation is shown in Figure~\ref{fig:chdem}. As mentioned above, we used three methods to derive the DEM and these are shown by different line styles in the figure. All of the DEM results suggest that the observed emission comes from mainly two different temperatures, a low temperature contribution centered at $\log T[\mathrm{K}] = 5.97$ and a high temperature contribution at $\log T[\mathrm{K}] = 6.20$. This is typical for coronal holes and we found similar results in the DEM analysis for all three locations we studied in the CH-B observation \citep[see also][]{Hahn:ApJ:2011}. We suspect that the cooler emission comes from the coronal hole itself, while the warmer component is from quiet Sun along the line of sight. 

\begin{figure}[h]
	\centering \includegraphics[width=0.9\textwidth]{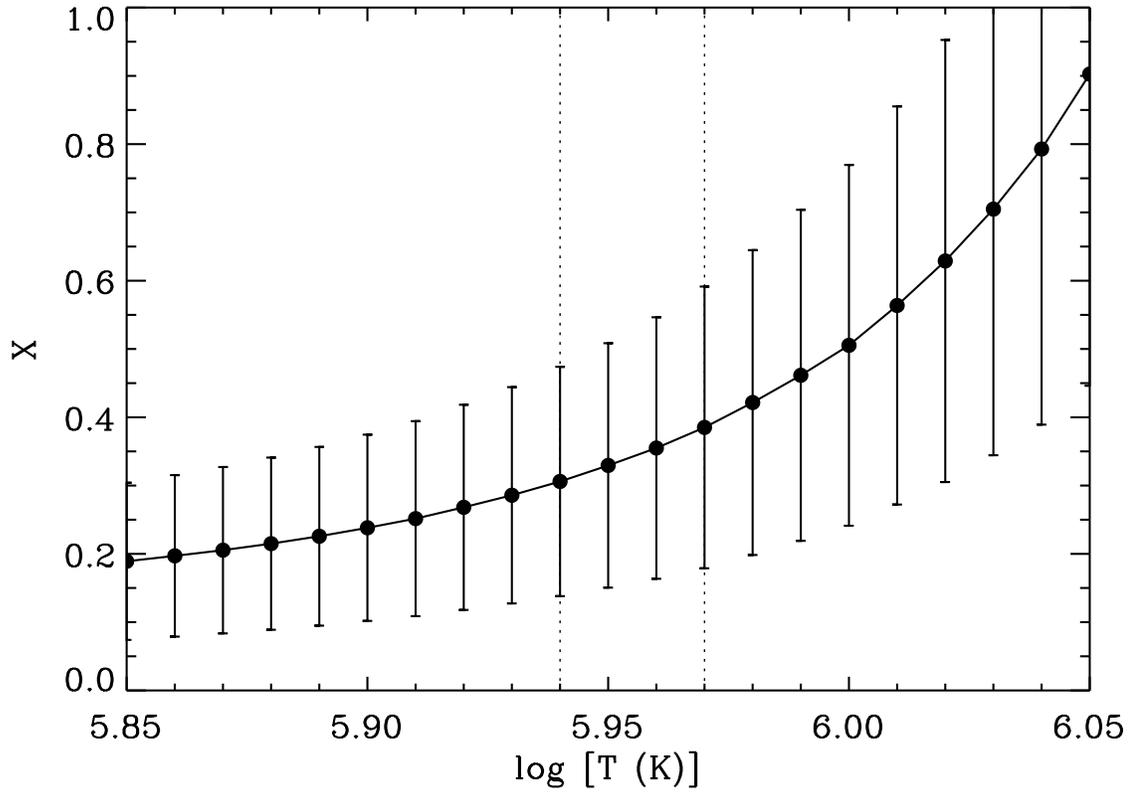}
	\caption{\label{fig:chirr} The irregularity $X$ as a function of the temperature $T$ for the CH-A observation at 1.05~$R_{\sun}$ using the Fe~\textsc{viii} diagnostic. Error bars indicate the statistical uncertainties from the intensity measurements that are propagated into the irregularity analysis. The vertical dotted lines indicate the temperatures most likely to be relevant: the formation temperature $\log T_t[\mathrm{K}] = 5.94$ or the peak DEM temperature at $\log T[\mathrm{K}] = 5.97$. 
}
\end{figure}

Figure~\ref{fig:chirr} shows the irregularity $X$ for the CH-A observation using the Fe~\textsc{viii} diagnostic as a function of the temperature $T$ assumed for the irregularity analysis. Based on our DEM results, the coronal hole material may be close to isothermal at a temperature of about $\log T[\mathrm{K}] = 5.97$, in which case we find $X = 0.39 \pm 0.21$. If the plasma is multithermal, the line formation temperature $\log T_t$ may be more relevant. For this observation Fe~\textsc{viii} has $\log T_{t}[\mathrm{K}] = 5.94$, which corresponds to $X = 0.31 \pm 0.17$. The Si~\textsc{vii} lines, which are formed at similar temperatures to Fe~\textsc{viii} show $X$ values of about $0.2$--$0.4$. In Section~\ref{subsec:slabinterpret} we show that these diagnostics are consistent given the uncertainties and considering their varying sensitivity to structure. 

Since the DEM indicates that there are two-temperature components, the Fe~\textsc{xii} and Fe~\textsc{xiii} lines that are formed at mainly in the higher temperature part of the DEM do not probe the same locations as the Si~\textsc{vii} and Fe~\textsc{viii} lines that are formed in the low temperature part of the DEM. The Fe~\textsc{xii} lines were bright enough to determine the irregularity in some of the coronal hole observations and we found values of $X \approx 0.1$ near the formation temperature of Fe~\textsc{xii}. These are smaller than we find for the same diagnostics in quiet Sun regions. This indicates that the Fe~\textsc{xii} emission in coronal holes is more structured than in quiet Sun regions, which is what we expect if the Fe~\textsc{xii} emission is coming from foreground or background material. In that case, the overall structure clearly differs from spherical symmetry because there is a curtain of quiet Sun along the line of sight through the coronal hole.

Focusing on the Fe~\textsc{viii} diagnostic, which is sensitive to coronal hole material and for which data exists across all the coronal hole observations, we find that plumes are more structured than interplumes relative to the assumed model. The CH-B interplume region has $X \sim 0.3$--$0.5$, consistent with the results from the interplume CH-A. However, the plume structure at $-10^{\prime\prime}$ in CH-B has $X\sim 0.05$ and the $90^{\prime\prime}$ plume has $X \sim 0.01$. For the plume at 90$^{\prime\prime}$, the density is relatively high $1\times 10^{9}$~$\mathrm{cm^{-3}}$, which is near upper limit of the Fe~\textsc{viii} density diagnostic and limits the accuracy of the irregularity diagnostic in that location. Nevertheless, since these results are all derived from the same diagnostic, it is valid to compare them and we can clearly see that the plumes are more structured than interplumes. Moreover, the plumes have quantitatively different amounts of structure from one another, which may be due to their overall transverse size or to differences in their internal structure. These results verify that our diagnostic is actually sensitive to structure, as a plume, by definition, is a structure within an interplume region. 

\subsection{Quiet Sun Region}\label{subsec:qs}

We also performed the irregularity analysis for a quiet Sun region. These data came from an EIS observation of the equatorial corona made on 2008 January 21. The EIS 1$^{\prime\prime}$ slit was a rastered across 22 horizontal positions with a step width of $10^{\prime\prime}$. More details about the observation can be found in \citet{Hahn:ApJ:2014}. Here, we binned the data by $8^{\prime\prime}$ in the vertical direction, but in the horizontal direction the bins are only one pixel ($1^{\prime\prime}$) wide. Since we are mainly interested in comparing general properties across regions we focused on a bin at $y=1.05$~$R_{\sun}$ located at solar coordinates (1030$^{\prime\prime}$, -23$^{\prime\prime}$). 

\begin{figure}[h]
	\centering \includegraphics[width=0.9\textwidth]{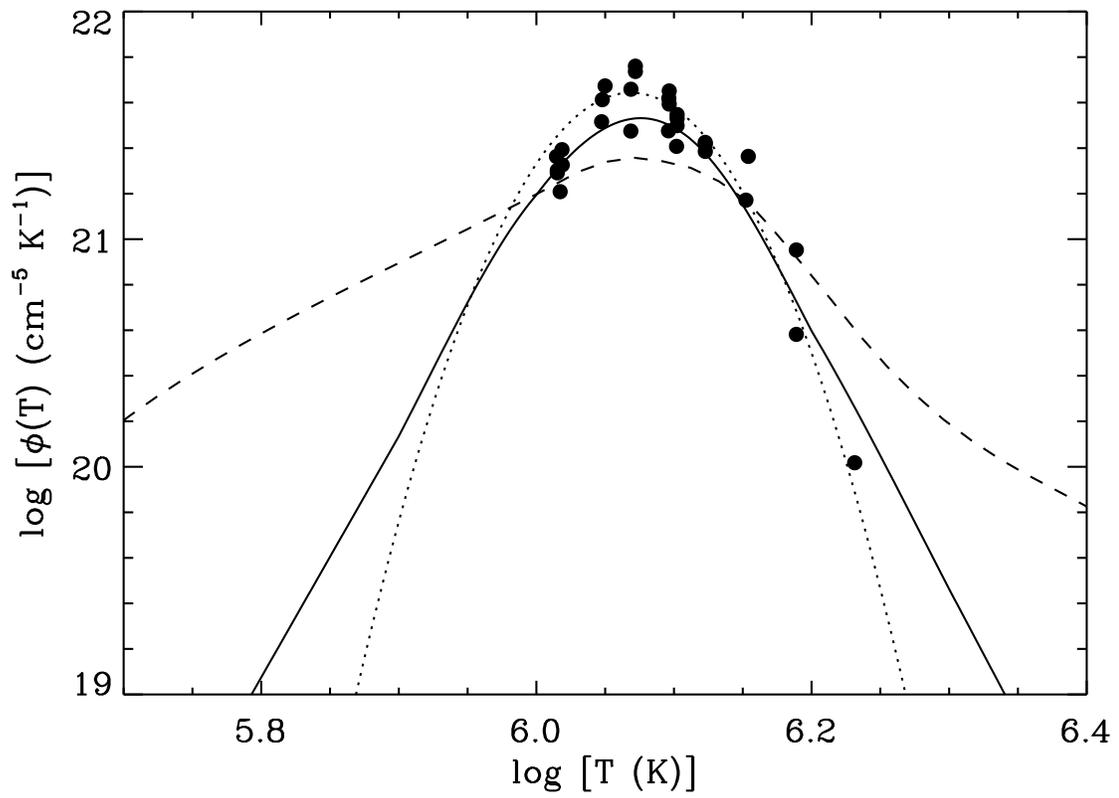}
	\caption{\label{fig:qsdem} Same as Figure~\ref{fig:chdem}, but for the quiet Sun observation
}
\end{figure}

Figure~\ref{fig:qsdem} shows the DEM results for the quiet Sun region. The iterative and regularized inversion methods both imply a relatively broad single-peaked temperature distribution centered around $\log T[\mathrm{K}] \approx 6.1$. A fit to a narrow, nearly isothermal, Gaussian function suggests a peak temperature somewhat higher, near $\log T[\mathrm{K}] = 6.17$. At low temperatures, the regularized inversion result seems to suggest a slower fall-off than the iterative method, but this likely a sysematic issue, because there is little data to constrain the DEM below about $\log T[\mathrm{K}] = 6.0$. 

\begin{figure}[h]
	\centering \includegraphics[width=0.9\textwidth]{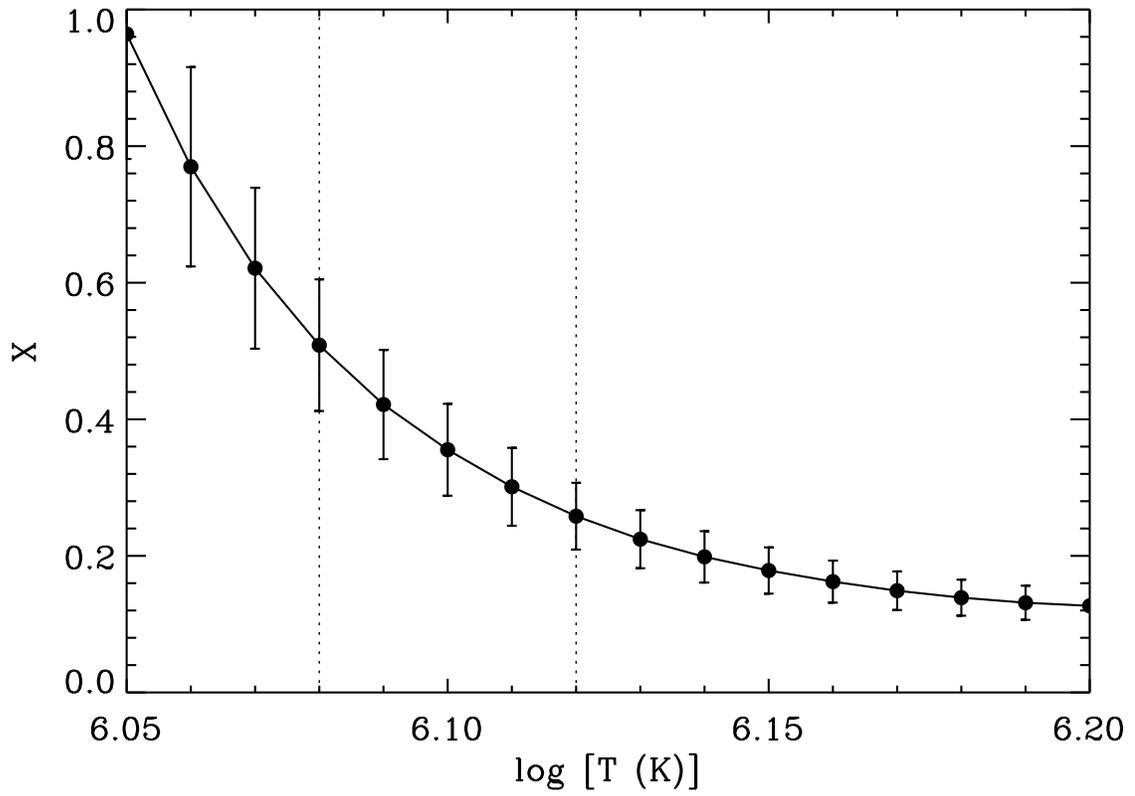}
	\caption{\label{fig:qsirr} Same as Figure~\ref{fig:chirr}, but for the quiet Sun observation and using the Fe~\textsc{xii} diagnostic. Here, the vertical dotted lines indicate the Fe~\textsc{xii} formation temperature of $\log T_t[\mathrm{K}]=6.12$ and the peak DEM for the Gaussian at $\log T[\mathrm{K}] = 6.08$.
}
\end{figure}

Irregularity results for the quiet Sun region are summarized in Table~\ref{table:irreg}. Figure~\ref{fig:qsirr} illustrates the sensitivity of $X$ on $T$ for the Fe~\textsc{xii} diagnostic. The quiet Sun has a temperature close to the formation temperature of the Fe~\textsc{xii} and Fe~\textsc{xiii} lines, making those the most useful diagnostics in this region. For Fe~\textsc{xii} at a formation temperature of $\log T_{t}[\mathrm{K}] = 6.12$, we find $X \approx 0.45$ and for Fe~\textsc{xiii} at $\log T_{t}[\mathrm{K}] = 6.15$, $X$ is about $0.41$. These roughly similar irregularities also agree when interpreted through in terms of a filling factor and density contrast, as discussed below in Section~\ref{sec:interpret}. If quiet Sun is isothermal at the peak temperature of the DEM, $\log T[\mathrm{K}] = 6.08$, then the Fe~\textsc{xii} and Fe~\textsc{xiii} diagnostics disagree, giving $X = 0.45$ and $X=2.3$, respectively. However, this is a rather low temperature for a quiet Sun region, where one more typically finds $\log T[\mathrm{K}] = 6.15$ \citep[][]{Phillips:Book}. This suggests that the formation temperature is more accurate in this case.

\subsection{Systematic Uncertainties}\label{subsec:uncertainties}

The major sources of systematic uncertainty in these results are the absolute intensity calibration of EIS and the uncertainties in the atomic data. The EIS intensity calibration has been drifting since \textit{Hinode} was launched in 2006. The original EIS calibration was performed prior to launch and had an estimated accuracy of about 20\% \citep{Lang:ApplOpt:2006}. Comparison of the EIS intensities with EUNIS observations in 2007 showed that EIS underestimated the intensities by about 20\%, which was consistent with EIS calibration uncertainties \citep{Wang:ApJS:2011}. A 20\% decrease since launch corresponds to a decay timescale of 5.6 years, and an exponential decay correction with a similar time constant of 1894 days was incorporated into the EIS data reduction software. The evolution of the EIS sensitivity has also been monitored through quiet Sun observations. \citet{Mariska:SolPhys:2013} found that the intensities of various lines apparently decayed over time, with a time constant of about 20 years, suggesting a very stable calibration. 

\citet{DelZanna:AA:2013}, however, studied a number of line intensity ratios that should be insensitive to either temperature or density. \citeauthor{DelZanna:AA:2013} found that a number of discrepancies among line intensity ratios measured by EIS compared with those measured by other instruments and what is predicted by atomic calculations. They proposed a correction to the EIS calibration that is a function of both time and wavelength. The revised calibration removes discrepancies among the line intensity ratios and also corrects some of the intensity drifts over time.

\citet{Warren:ApJS:2014} also proposed a revised calibration, which they obtained by comparing full-Sun EIS mosaics to EVE data. The absolute calibration of EVE is believed to be reliable as it is checked periodically by rocket measurements. The correction proposed by \citeauthor{Warren:ApJS:2014} is also a complex function of time and wavelength. Because EVE data is only available since 2010, the calibrations for earlier dates are only an extrapolation. Nevertheless, the calibration does successfully correct many of the drifts found by the EIS synoptic monitoring, including some that are not completely removed by the \citeauthor{DelZanna:AA:2013} calibration. In particular, the drift in the Fe~\textsc{viii} 185.21~\AA\ intensity is corrected, which is an important line for our diagnostics as it is sensitive to low densities. One possible problem with the \citeauthor{Warren:ApJS:2014} calibration is that it suggests the EIS to EUNIS calibration discrepancy is a factor of two rather than 20\%. The source of this discrepancy is unclear.

Based on tests using both methods, we have chosen to use the \citeauthor{Warren:ApJS:2014} calibration. We performed a DEM analysis of the various observations using both proposed calibrations. The \citeauthor{Warren:ApJS:2014} calibration almost always produced DEM results with smaller chi-squared values than those found using the \citeauthor{DelZanna:AA:2013} calibration. This suggests that that calibration is more consistent for comparing line intensities at different wavelengths.

\begin{figure}[h]
	\centering \includegraphics[width=0.9\textwidth]{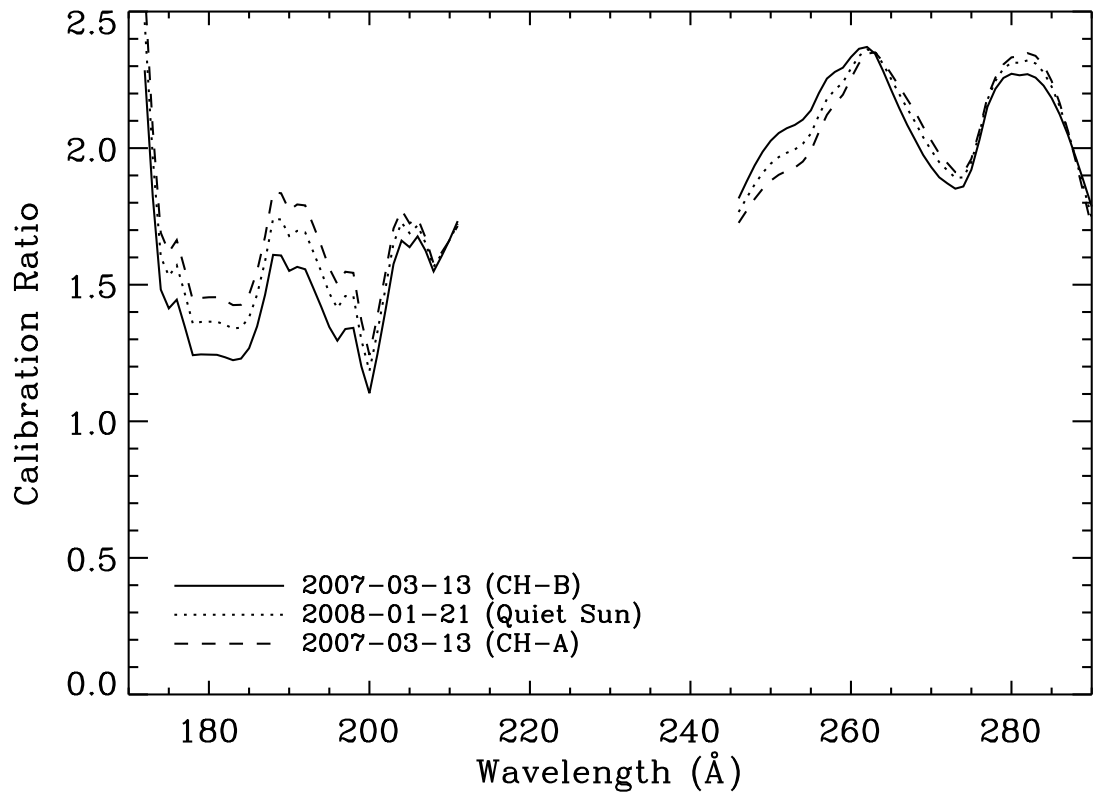}
	\caption{\label{fig:calcompare} Ratio of the absolute calibration using the \citet{DelZanna:AA:2013} method compared to the one from \citet{Warren:ApJS:2014} as a function of wavelength. The various line styles correspond to the dates of the three observations we studied. 
}
\end{figure}

The irregularity factors are proportional to the absolute intensities. Thus, we can estimate the effects the different calibrations can have on our inferred $X$ valuess. Figure~\ref{fig:calcompare} shows the ratios of the \citet{DelZanna:AA:2013} calibration to the \citet{Warren:ApJS:2014} calibration. The \citeauthor{DelZanna:AA:2013} calibration implies intensities larger by about 50\% in the EIS short wavelength band and about a factor of two larger in the long wavelength band. Multiplying our $X$ results by the appropriate factor, gives an estimate of how our results would change by using the \citet{DelZanna:AA:2013} calibration. (Though, this is a rough estimate as it ignores the changes in the relative calibration, which also affect $X$ through the line ratio). Because, the \citeauthor{Warren:ApJS:2014} calibration implies smaller $X$ values, we infer a greater absolute degree of structure than we would if we had used the \citeauthor{DelZanna:AA:2013} calibration. However, tests with both cases do find the same general trends, such as plume regions having smaller $X$ than interplumes. Additionally, we found that the inferred $X$ values for Fe~\textsc{xii} and \textsc{xiii} in the quiet Sun are nearly the same using either calibration. 

An additional source of systematic uncertainty is the atomic data. In particular, there can be large uncertainties in the interpretation of density diagnostics. \citet{Young:AA:2009}, for example, found discrepancies of up to a factor of three between the Fe~\textsc{xii} and Fe~\textsc{xiii} density diagnostics. Updates to the atomic data may have reduced the disagreement \citep{DelZanna:AA:2012, DelZanna:AA:2012b}. However, uncertainties in these and other line ratios remain. These uncertainties are difficult to quantify, since most atomic data results are based on theoretical calculations that do not provide a clear estimate of their uncertainties. Experimental studies to benchmark the atomic calculations are needed.

\section{Interpretation Using a Two-Density Slab Model}\label{sec:interpret}

The irregularity parameter itself does not give a physical description of the density structure. Rather, it measures the relative level of unresolved structure compared to an assumed large-scale density distribution, in this case, spherical symmetry. In order to relate $X$ to physical quantities, a model for the unresolved structure is needed. Here we describe a simple two-density slab model. 
	
For the model, we consider the density distribution along the line of sight to take only two values, which we label $n_{\mathrm{A}}$ and $n_{\mathrm{B}}$. We do not consider any other density variations along the line of sight.
For the slab model, we define $\alpha$ to be the fraction of the path along the line of sight that has density $n_{\mathrm{A}}$ so that $(1-\alpha)$ of the material has $n_{\mathrm{B}}$. The observed intensity of a density-sensitive line $I_{1}$ will be
\begin{equation}
I_{1} \propto \alpha \epsilon_{1} (n_{\mathrm{A}}) + (1-\alpha) \epsilon_{1} (n_{\mathrm{B}}).
\label{eq:i1prop}
\end{equation}
The emissivities are calculated for a fixed temperature, the choice of which depends on the context of the model.

Given two lines, $I_1$ and $I_2$, that form a density diagnostic, their ratio is 
\begin{equation}
\frac{I_{1}}{I_{2}} = \frac{\alpha \epsilon_{1} (n_{\mathrm{A}}) + (1-\alpha) \epsilon_{1} (n_{\mathrm{B}})}
{\alpha \epsilon_{2} (n_{\mathrm{A}}) + (1-\alpha) \epsilon_{2} (n_{\mathrm{B}})}.
\label{eq:rat12}
\end{equation}
This ratio can be interpreted as the one that would result if the slab were uniform with density $n_{0}$:
\begin{equation}
\frac{I_{1}}{I_{2}} = \frac{ \epsilon_{1}(n_{0})}{\epsilon_{2}(n_{0})}. 
\label{eq:defn0}
\end{equation}

For a slab, we again define the irregularity as the ratio of the observed intensities to the one that would result if the density were uniform at $n_{0}$ along the line of sight. That is, 
\begin{equation}
X = \frac{\alpha \epsilon_{1} (n_{\mathrm{A}}) + (1-\alpha) \epsilon_{1} (n_{\mathrm{B}})}{\epsilon_{1} (n_{0})}
=\frac{\alpha \epsilon_{2} (n_{\mathrm{A}}) + (1-\alpha) \epsilon_{2} (n_{\mathrm{B}})}{\epsilon_{2} (n_{0})}.
\label{eq:slabxdef}
\end{equation}
The irregularities obtained from the two line intensities are the same. 

The slab model has three parameters, $n_{\mathrm{A}}$, $n_{\mathrm{B}}$, and $\alpha$ (from which we can derive $n_{0}$). However, it is convenient to reduce this to two quantities that are often discussed. Although $n_{\mathrm{A}}$ and $n_{\mathrm{B}}$ are interchangeable, we can interpret the greater of the two to represent the structure and the lesser to represent the background. Then, we define the filling factor $f$ to be the fraction along the line of sight that has that greater density. This corresponds to $\alpha$ if $n_{\mathrm{A}} > n_{\mathrm{B}}$ and $1-\alpha$ if $n_{\mathrm{A}} < n_{\mathrm{B}}$. We also define the density contrast, $q$, to refer to the ratio of the larger density relative to the smaller one, so that $q \geq 1$. 

\subsection{Properties of the Slab Model}\label{subsec:slabgeneral}

Below, we interpret our irregularity results from Section~\ref{sec:obs} using the slab model. But first it is useful to examine the general behavior of the irregularity diagnostics for a given set of slab model parameters. To illustrate this, we pick a few parameters and then use the model to calculate the resulting irregularity. 

In order to define the model, we need a set of parameters $n_{\mathrm{A}}$, $n_{\mathrm{B}}$, and $\alpha$. However, since observations infer the average density $n_{0}$, it is more useful to consider that as a parameter than $\alpha$. If $n_{\mathrm{A}}$, $n_{\mathrm{B}}$, and $n_{\mathrm{0}}$ are all given, then the corresponding $\alpha$ can be determined algebraically from Equations~(\ref{eq:rat12}) and (\ref{eq:defn0}). For the example described here, we set $n_{0} = 1 \times 10^{8}$~$\mathrm{cm^{-3}}$. We set $n_{\mathrm{A}} = 2 \times 10^{7}$, to represent a low density background. Finally, we choose various values for $n_{\mathrm{B}}$ ranging from $n_{0}$ to $1 \times 10^{9}$~$\mathrm{cm^{-3}}$. We set $\log T[\mathrm{K}]$ = 6 for calculating all the emissivities. 

\begin{figure}[h]
	\centering \includegraphics[width=0.9\textwidth]{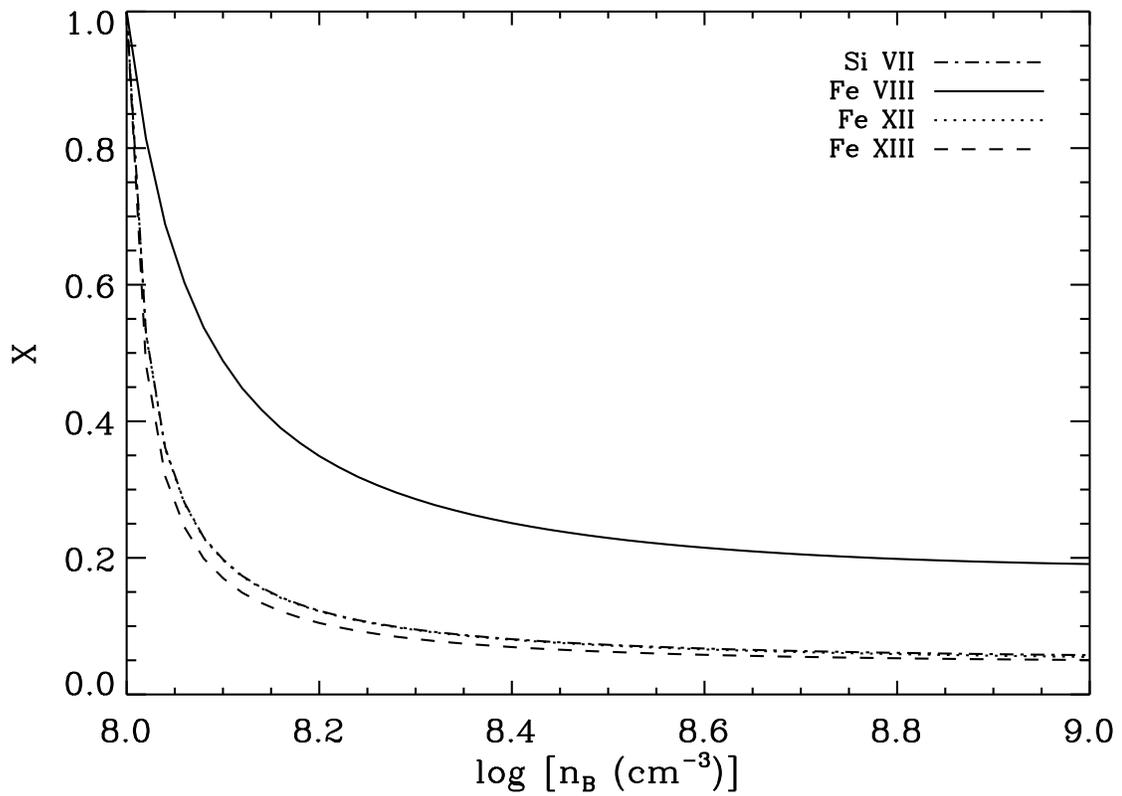}
	\caption{\label{fig:slab1} The inferred irregularity $X$ as a function of the $n_{\mathrm{B}}$ for a slab model where $n_{0} = 1 \times 10^{8}$~$\mathrm{cm^{-3}}$ and $n_{\mathrm{A}} = 2 \times 10^{7}$~$\mathrm{cm^{-3}}$. The different line styles refer to the irregularity diagnostics we have used, labeled according to ion. The Si~\textsc{vii} and Fe~\textsc{xii} diagnostics have a very similar response of $X$ versus $n_{\mathrm{B}}$ for the parameters chosen here and overlap in the figure.
}
\end{figure}

Figure~\ref{fig:slab1} shows the inferred irregularity as a function of $n_{\mathrm{B}}$ for the various diagnostic line ratios we used in our analysis. This shows that $X$ varies inversely with the density contrast, $n_{\mathrm{B}}/n_{\mathrm{A}}$. For $\log [n_{\mathrm{B}}]= 8.0$, we find $X=1$, because in this case $n_{\mathrm{B}}=n_{\mathrm{0}}$ and there is no structure. For the same situation, $\alpha = 0$, because $\alpha$ corresponds to the fraction of the emission coming from the low density structure at $\log [n_{\mathrm{A}}] = 7.3$. The figure also shows that $X$ tends to saturate at density contrasts above about a factor of ten (corresponding to $\log[ n_{\mathrm{B}}] = 8.3$). 

\begin{figure}[h]
	\centering \includegraphics[width=0.9\textwidth]{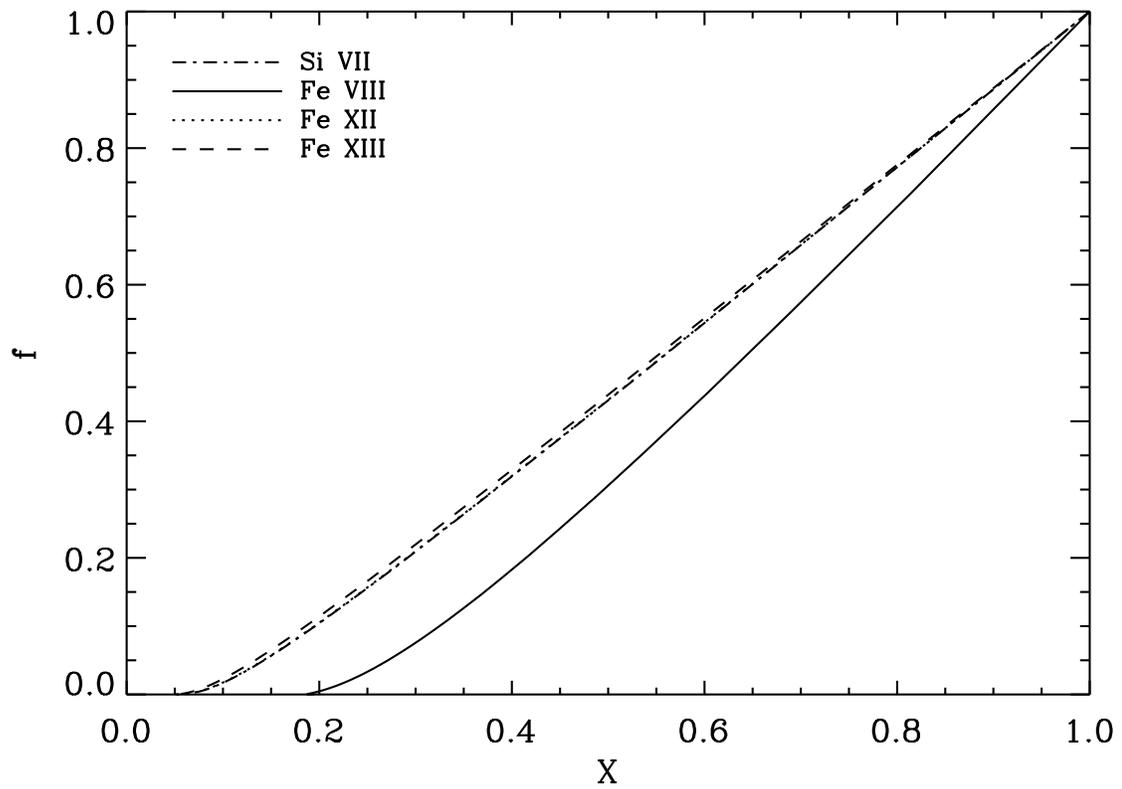}
	\caption{\label{fig:slab2} Same as Figure~\ref{fig:slab1}, but showing the filling factor, $f$, as a function of the irregularity $X$, for various diagnostic line ratios. There is a strong correlation between $f$ and $X$, but the absolute relation differs among the various diagnostics. 
}
\end{figure}

Figure~\ref{fig:slab2} illustrates the relation between the filling factor, $f$, and $X$ for the same set of model parameters as viewed above. Here, the filling factor is $f=1-\alpha$, corresponding to the fraction of emission coming from density $n_{\mathrm{B}}$, since we have set $n_{\mathrm{B}} > n_{\mathrm{A}}$ in every case. It is clear that the filling factor and the irregularity are strongly correlated.

The above results demonstrate the strong correlations among $X$ and physical parameters, such as filling factor and density contrast. However, they also show that the different diagnostic lines all have varying responses to these parameters. Thus, $X$ from a given diagnostic can be used consistently to quantify the relative level of structure in various observations. But a model is needed in order to quantify the structure or to compare $X$ values from different diagnostics. 

\subsection{Application to Observed Irregularities}\label{subsec:slabinterpret}

We have used the slab model to find the density contrasts and filling factors that are consistent with our observations. From our measurements, we determined the temperature, the line intensity ratio, and the irregularity parameter. The observed temperature defines the value to be used to calculate the emissivities and the observed intensity ratio defines the value of $n_{0}$ to be used in the model. Having fixed the value of $n_{0}$, we then find values of $n_{\mathrm{A}}$ and $n_{\mathrm{B}}$ that produce the observed $X$. If all three densities, $n_{0}$, $n_{\mathrm{A}}$, and $n_{\mathrm{B}}$ are given, then $\alpha$ can be found algebraically. The procedure we use is to first pick an arbitrary value for $n_{\mathrm{A}}$ and then numerically solve for the $n_{\mathrm{B}}$ and $\alpha$ that produce the observed irregularity $X$, while simultaneously yielding the intensity ratio $I_1/I_2$ that corresponds to $n_{0}$. Since we have essentially two equations with three unknowns, our results define a curve in parameter space giving $n_{\mathrm{B}}$ as a function of $n_{\mathrm{A}}$. As mentioned above, we change variables to report this more usefully as the filling factor $f$ and density contrast $q$. 

\begin{figure}[h]
	\centering \includegraphics[width=0.9\textwidth]{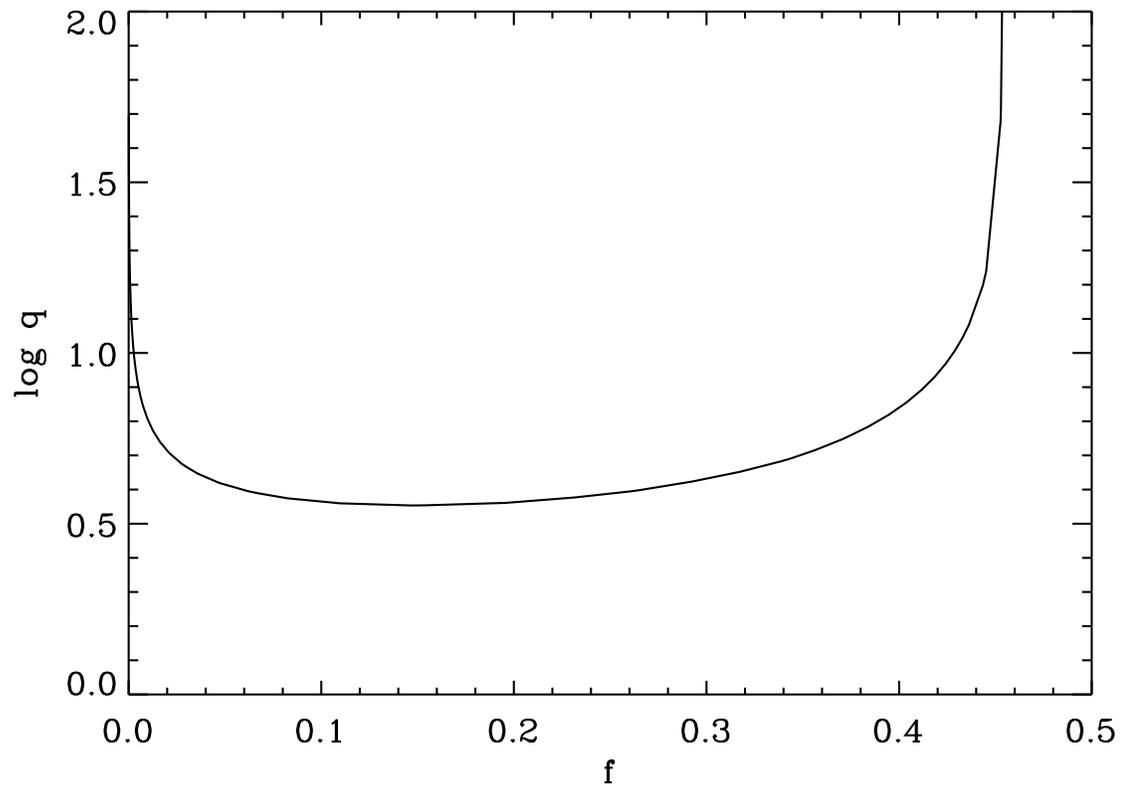}
	\caption{\label{fig:slabinterp} The curve indicates the set of $\log q$ and $f$ values that are consistent with the observed results for the quiet Sun observation obtained using the Fe~\textsc{xii} diagnostic. Every $q$ and $f$ on the curve input to the slab model, will reproduce the observed average density of $n_{\mathrm{0}}=3.0 \times 10^{8}$~$\mathrm{cm^{-3}}$ and irregularity $X = 0.45$. 
}
\end{figure}

Figure~\ref{fig:slabinterp} plots the set of $f$ and $q$ that are consistent with the quiet Sun observation for Fe~\textsc{xii}, which inferred $X=0.45$ and an average density of $n_{\mathrm{0}} = 3.0 \times 10^{8}$~$\mathrm{cm^{-3}}$. The various properties of this plot can be explained as follows: At the left end of the plot, there are very small filling factors with large density contrasts; this corresponds to a situation where there are very small, but extremely dense structures embedded in a plasma that is otherwise at a density close to but somewhat less than $n_{0}$. On the right side of the plot, there are also very large density contrasts, but with moderately large filling factors near $\sim 0.5$; here, the denser component has a density close to $n_{0}$, but the less dense component is much less dense, as if there were large voids in the plasma. Intermediate values of $f$ and $q$ fall between these two extreme situations. All of our results for other observations and diagnostic ratios share these qualitative properties, though they differ quantitatively. 

In order to infer a single value for $f$ and $q$ we choose the minimum $q$ and the corresponding $f$ from the slab model. In Figure~\ref{fig:slabinterp}, this corresponds to $q = 3.5$ ($\log q = 0.55$), and $f=0.15$. This seems reasonable, for several reasons. First, this represents a lower bound on the density contrast $q$ (leaving aside the uncertainties). Second, a moderate value of $f$ seems more plausible than either of the extreme situations with large $q$. That is, if there were tiny structures that had densities orders of magnitude larger than the background, or if there were essentially empty voids in the plasma, we would expect them to have other signatures and we would not have to resort to complex methods to indirectly infer their existence. For example, they might have signatures in the solar wind that would be detected by in-situ observations.

The minimum $q$ and corresponding $f$ for each observation are summarized in Table~\ref{table:irreg}. For interplume regions of coronal holes and quiet Sun regions, we find minimum density contrasts of a factor of $\sim 3$--10, with corresponding filling factors of $\sim10$--20\%. For plume regions of coronal holes, where we can see that there is obvious structure, we find much smaller filling factors of only a few percent and very large density contrasts of more than a factor of ten. Such large density contrasts in the plume regions are surprising. However, note that Figure~\ref{fig:slab1} shows that at large density contrasts $X$ saturates, so that very small errors in $X$ can lead to large differences in the inferred $q$.

Considering possible errors in the EIS absolute calibration, these results would be significantly different if the \citet{DelZanna:AA:2013} calibration correction had been used. Since the effect of that calibration would be to roughly double the absolute intensities, we would also find values of $X$ about a factor of two larger. Since $f$ is roughly proportional to $X$, this calibration also doubles the filling factor $f$ and reduces the density contrast. 

In some cases, we find values for $X$ that are greater than one. These values cannot be reproduced using the slab model. Such results may be due errors the temperature input to the irregularity analysis, or to errors in the atomic data. However, there are also physical reasons that $X$ might be greater than one, as we discuss in the next section. 

\subsection{Limitations of the Slab Model}\label{subsec:slablimit}

The slab model has several obvious limitations. First, we are assuming that there are only two densities along the line of sight, whereas in reality there is a distribution of densities. Even if there are only two types of underlying structures with two characteristic densities, these densities would be smeared out along the line of sight due to the radial density fall-off with height. Because structures are distributed on the Sun, the line of sight would pass through the structures at different radial heights. A more complex model could introduce a distribution of densities, but this would expand the parameter space, thereby requiring additional observational or theoretical constraints. 

Another limitation of the model is that it is isothermal. The assumption that the observed region is isothermal is also implicit in the irregularity analysis itself. However, this is not a major limitation of the irregularity analysis, since the assumption can be thought of as a normalization. The irregularity measures the deviation of the observation from an isothermal spherically symmetric corona. The interpretation of this deviation in terms of a physical quantities is left to the slab model. The DEM analysis suggests that an isothermal assumption is reasonable. All of the DEM results show either a single clear peak or two peaks that correspond to typical coronal hole and quiet Sun temperatures. 

The slab model could be extended to allow the two densities to also have different temperatures. Then, for example, we would replace $\epsilon_{1}(T_{0}, n_{A})$ by $\epsilon_{1}(T_{\mathrm{A}}, n_{A})$, where $T_{0}$ is the observed temperature currently used in the model and $T_{\mathrm{A}}$ is the temperature of the structure that has density $n_{\mathrm{A}}$. Similar substitutions would be made for the other quantities in the slab model analysis. We have tested this, and these changes do allow us to reproduce $X$ values greater than one. However, this happens at the expense of introducing new parameters and requiring new constraints. At this point the slab model becomes a crude analysis for the emission measure differential in both temperature and density. Such an analysis would clearly be more complete than what we have done here, but it is not currently practical, as was discussed in Section~\ref{sec:intro}. 

\section{Discussion and Conclusions}\label{sec:discuss}

The irregularity analysis reveals that coronal holes and quiet Sun regions have density contrasts of at least a factor of 3--10 and filling factors on the order of 10--20\%. These results are in rough agreement with what has been found using other methods. For example, \citet{Raymond:ApJ:2014} found density contrasts of at least a factor of six, although they inferred a somewhat larger filling factor of about 50\%. 

We have also found that the irregularity parameter successfully measures the relative amount of unresolved density structure in observations. For example, plumes are more structured than interplumes. These relative measurements are fairly robust to the approximations required by the analysis, and are independent of the interpretation of $X$ in terms of physical quantities. 

Our method offers several advantages compared to other methods that have been used to infer properties of the unresolved density structure. A common method to determine the filling factor is to compare the density and the emission measure, but to derive $f$ from those quantities requires an estimate of the length scale of the structure. For some structures, such as an active region loop, the needed length scale can be obtained from images. But for structures like a coronal hole or quiet Sun region, it is not clear what length scale should be used. The irregularity analysis resolves this problem by considering deviations from an assumed large scale density distribution, which we have chosen to be a spherically symmetric scale-height falloff. However, the analysis could be refined in the future by using a more detailed representation of the large-scale structures of the corona. 

The irregularity analysis described here is based solely on EUV data. This removes possible problems with the relative calibration between instruments. It also allows a variety of diagnostics that are sensitive to different temperatures to be used, which helps to distinguish foreground structures in analysis. For example, to study a coronal hole, we can use diagnostic lines formed at cool temperatures and thereby avoid contributions from warmer quiet Sun material along the line of sight.

The most significant limitation for the current analysis is the factor of two uncertainty in the EIS absolute calibration. This affects the interpretation of our results in terms of filling factor and density contrast, although the relative irregularities among different regions are fairly robust to these errors. The analysis can be improved in the future by using rocket measurements, which are expected to have a more accurate absolute calibration. Such measurements, will also provide new data that may lead to a more definitive calibration for EIS \citep[][]{Wang:ApJS:2011, Warren:ApJS:2014}. 

The uncertainties in the atomic data are a significant limitation, which also affect all other solar analyses that incorporate density. The resulting uncertainties are difficult to quantify, because most of the data come from theoretical calculations and experimental tests of those calculations are lacking. Density diagnostics seem particularly sensitive to these errors, because the line intensities depend on all of the processes that determine the level populations. This is in contrast to temperature diagnostics, for example, where the sensitivity is dominated by the ionization balance. Density diagnostics also depend on taking a line ratio, which magnifies the uncertainties of either line individually. Laboratory measurements are needed in order to assess the uncertainties in the available atomic data.
	
\begin{acknowledgments}
This work was supported in part by the NASA Living with a Star program grant NNX15AB71G and by the NSF Atmospheric and Geospace Sciences Solar Heliospheric and Interplanetary Environment program grant 1459247. 
\end{acknowledgments}

\clearpage
		
\bibliography{structure}

\newpage

\renewcommand{\arraystretch}{0.50}
\begin{deluxetable}{llccrcl}
	\tabletypesize{\footnotesize}
	\tablecaption{Transitions for Density Diagnostics. \label{table:densitylines}}
	\tablewidth{0pt}
	\tablehead{
		\colhead{Pair} & 
		\colhead{Ion} & 
		\colhead{~} & 
		\colhead{$\lambda$ (\AA)\tablenotemark{a}} & 
		\multicolumn{3}{c}{Transition\tablenotemark{a}}
		}
	\tablecolumns{6}
	\startdata
1	&Si \textsc{vii} & &274.18 & $ 2s^2\, 2p^4\, ^{3}P_1 $ &$-$ & $2s\, 2p^5\, ^{3}P_0 $ \\
1	&Si \textsc{vii} & &275.36 & $ 2s^2\, 2p^4\, ^{3}P_2 $ &$-$ & $2s\, 2p^5\, ^{3}P_2 $ \\
2	&Fe \textsc{viii} & &185.21 & $ 3p^6\, 3d\, ^{2}D_{5/2} $ &$-$ & $3p^5\,3d^2\, (^{3}F)\, ^{2}F_{7/2}$ \\
2	&Fe \textsc{viii} & & 186.60 & $3p^6 \,3d\, ^{2}D_{3/2} $ &$-$ & $3p^5\, 3d^2\, (^{3}F)\, ^{2}F_{5/2}$ \\
	& & &186.85 & $ 3s^2\, 3p^3\, ^{2}D_{3/2}$ & $-$ & $3s^2\, 3p^2\, 3d\, ^{2}F_{5/2}$ \\
\raisebox{1.5ex}[0pt]{3}	&\raisebox{1.5ex}[0pt]{Fe \textsc{xii}} & \raisebox{1.5ex}[0pt]{$\Big\{$} &186.89 & $ 3s^2\, 3p^3\, ^{2}D_{5/2}$ & $-$ & $3s^2\, 3p^2\, 3d\, ^{2}F_{7/2}$ \\
3	&Fe \textsc{xii} & &195.12 & $ 3s^2\, 3p^3\, ^{4}S_{3/2} $ &$-$ & $3s^2\, 3p^2\, (^{3}P)\, 3d\, ^{4}P_{5/2} $ \\
4	&Fe \textsc{xiii} & &202.04 & $ 3s^2\, 3p^2\, ^{3}P_0 $ &$-$ & $3s^2\, 3p\, 3d\, ^{3}P_1 $ \\
	&  & &203.80 & $ 3s^2\, 3p^2\, ^{3}P_{2}$ &$-$ & $ 3s^2\, 3p\, 3d\, ^{3}D_{2}$ \\*
	&  & &203.83 & $ 3s^2\, 3p^2\, ^{3}P_{2}$ &$-$ & $ 3s^2\, 3p\, 3d\, ^{3}D_{3}$ \\*
\raisebox{3.5ex}[0pt]{4} &\raisebox{3.5ex}[0pt]{Fe \textsc{xiii}} & \raisebox{3.5ex}[0pt]{$\Bigg\{$} &203.84 & $ 3s^2\, 3p^2\, ^{3}D_{2}$ &$-$ & $3s\, 3p^2\, 3d\, ^{3}F_{2} $
	\enddata
	\tablenotetext{a}{Wavelengths and transitions taken from CHIANTI \citep{Dere:AA:1997, DelZanna:AA:2015}.}
	\tablecomments{The ``pair'' column labels the transitions that when used together form a density diagnostic. Brackets indicate blends from the same ion.}
\end{deluxetable}

\newpage
\begin{deluxetable}{lllCCCC}
	\tabletypesize{\footnotesize}
	\tablecaption{Irregularity Results. \label{table:irreg}}
	\tablewidth{0pt}
	\tablehead{
		\colhead{Region} & 
		\colhead{Diagnostic} & 
		\colhead{Temperature} & 
		\colhead{$\log T[\mathrm{K}]$} & 
		\colhead{$X$} &
		\colhead{$\log q$} & 
		\colhead{$f$}
	}
	\tablecolumns{6}
	\startdata
	CH-A & Si~\textsc{vii}  & $T_t$ & 5.94 & 0.26 \pm 0.53 		& 0.7 & 0.1  \\ 
	CH-A & Si~\textsc{vii}  & peak & 5.97 & 0.36 \pm 0.73 & 0.6 	& 0.1 \\ 
	CH-A & Fe~\textsc{viii} & $T_t$ & 5.94 & 0.31 \pm 0.17 		& 1 & 0.1 	\\ 
	CH-A & Fe~\textsc{viii} & peak &  5.97 & 0.39 \pm 0.21& 0.9	& 0.1 \\ 
	CH-B Interplume at -50$^{\prime\prime}$ & Si~\textsc{vii}  & $T_t$ & 5.96 		& 0.6 \pm 3.3 	& 0.5 & 0.2 \\
	CH-B Interplume at -50$^{\prime\prime}$ & Si~\textsc{vii}  & peak & 6.00 	& 1.0 \pm 5.3 	& \nodata & \nodata \\
	CH-B Interplume at -50$^{\prime\prime}$	& Fe~\textsc{viii} & $T_t$ & 5.96 		& 0.33 \pm 0.28 & 0.9 & 0.1 \\
	CH-B Interplume at -50$^{\prime\prime}$	& Fe~\textsc{viii} & peak & 6.00 	& 0.47 \pm 0.38 & 0.8 & 0.2 \\  
	CH-B Plume at -10$^{\prime\prime}$	& Fe~\textsc{viii} & $T_t$ & 5.94 & 0.05 \pm 0.12 			& 1.8 & 0.02 \\ 
	CH-B Plume at -10$^{\prime\prime}$	& Fe~\textsc{viii} & peak & 5.97 & 0.07 \pm 0.15 		& 1.7 & 0.02 \\ 
	CH-B Plume at 90$^{\prime\prime}$ & Fe~\textsc{viii} & $T_t$ & 5.96 & 0.003 \pm 0.065 			& 3.1 & 0.002 \\ 
	CH-B Plume at 90$^{\prime\prime}$ & Fe~\textsc{viii} & peak & 6.00 & 0.007 \pm 0.092 		& 2.7 & 0.004 \\ 
	QS  & Fe~\textsc{xii} & $T_t$ & 6.12 & 0.23 \pm 0.04 		& 0.8 & 0.1 \\
	QS  & Fe~\textsc{xii} & peak & 6.08 & 0.45 \pm 0.08 	& 0.6 & 0.1 \\
	QS  & Fe~\textsc{xiii} & $T_t$ & 6.15 & 0.41 \pm 0.11 		& 0.6 & 0.1 \\
	QS  & Fe~\textsc{xiii} & peak & 6.08 & 2.3 \pm 0.6 	& \nodata & \nodata
	\enddata
	\tablecomments{Temperatures correspond to either the formation temperature for the line $T_t$ or the peak temperature of the DEM. The uncertainties on $X$ reported here are statistical and do not include errors from absolute calibration and other systematic sources. The $f$ and $q$ values estimate the minimum $q$ and the corresponding $f$ consistent with $X$, ignoring the uncertainties in $X$.}
\end{deluxetable}

\end{document}